\documentclass[conference, letterpaper]{IEEEtran}
\usepackage{xcolor}
\usepackage{fancyhdr}
\usepackage{multirow}
\usepackage{amsmath,amssymb,amsfonts,mathabx}
\usepackage{titlesec}
\usepackage{rotating}
\usepackage{tikz}
\usepackage{multirow}
\usepackage{mathrsfs}
\usepackage{amsthm}
\usepackage{amssymb}
\usepackage{cite}
\usepackage{mdframed}
\usepackage{braket}
\usepackage{stmaryrd}
\usepackage{mathtools}
\usepackage{comment}
\usepackage{caption}
\usepackage{subcaption}
\theoremstyle{definition}
\newtheorem{definition}{Definition}

\def\reals{\mathbb{R}}

\def\nats{\mathbb{N}}

\def\xb{x}
\def\ub{u}
\def\etab{\boldsymbol\eta}
\def\xbb{\bar{\boldsymbol x}}
\def\ubb{\bar{\boldsymbol u}}
\def\Xb{\bar{X}}
\def\Ub{\bar{U}}

\DeclareMathSymbol{\mlq}{\mathord}{operators}{``}
\DeclareMathSymbol{\mrq}{\mathord}{operators}{`'}

\newcommand{\ball}{\Omega}
\newcommand{\Sol}{\mathit{Sol}}

\newcommand{\LTLalways}{\Box}
\newcommand{\LTLeventually}{\Diamond}
\newcommand{\LTLuntil}{\mathbin{\sf U}}

\emergencystretch=\maxdimen
\ifCLASSINFOpdf

\else

\fi

\usepackage{amsmath}

\begin{document}

\title{Logic-based Knowledge Awareness for Autonomous Agents in Continuous Spaces}

\author{Arabinda Ghosh*, Mahmoud Salamati*, and
Sadegh Soudjani \thanks{All authors are with Max Planck Institute for Software Systems, Germany. {\tt\small \{arabinda, msalamati, sadegh\}@mpi-sws.org}.
\newline * indicates equal contributions. 
\newline
This work is supported by the Horizon Europe EIC project SymAware under the grant agreement 101070802.}}
\date{January 2024}
\newcommand{\lef}[1]{{\color{blue}[LV] #1}}

\IEEEoverridecommandlockouts

\maketitle
\thispagestyle{plain}
\pagestyle{plain}

\begin{abstract}
This paper presents a step towards a formal controller design method for autonomous agents based on knowledge awareness to improve decision-making. Our approach is to first create an organized repository of information (a knowledge base) for autonomous agents which can be accessed and then translated into temporal specifications. Secondly, to develop a controller with formal guarantees that meets a combination of mission-specific objective and the specification from the knowledge base, we utilize an abstraction-based controller design (ABCD) approach, capable of managing both nonlinear dynamics and temporal requirements. Unlike the conventional offline ABCD approach, our method dynamically updates the controller whenever the knowledge base prompts changes in the specifications. A three-dimensional nonlinear car model navigating an
urban road scenario with traffic signs and obstacles is considered for validation. Results show the effectiveness of the method in guiding the autonomous agents to the target while complying with the knowledge base and the mission-specific objective.  
\end{abstract}

\IEEEpeerreviewmaketitle

\section{Introduction}

 Autonomous agents are becoming popular in various application areas, including road driving, warehouse exploration, and space exploration~\cite{gehrke2008evaluating}. The agents in these systems use sensors like cameras, lidar, radar, and GPS to detect and perceive their surroundings. The data from these sensors, along with the map information, must be represented in a way that the autonomous agents can comprehend. Therefore, The main challenges in these systems are (1) to form a formal \emph{specification} by integrating the sensory data with the existing \emph{knowledge}, and (2) to synthesize a \emph{controller} that can provide \emph{formal guarantee} over satisfaction of the specification.
 
 Our approach to address the first challenge is to create an organized repository that would enable autonomous agents to efficiently access available information, and then translate the information into the form of a temporal specification. In this context, we explore the field of \emph{knowledge representation} techniques which support the construction of a \emph{knowledge base} (a centralized or decentralized repository)  that systematically stores information~\cite{fagerberg2012innovation}. \emph{Knowledge awareness}, on the other hand, refers to the ability of an intelligent system to understand, interpret, and utilize the information contained within the knowledge base~\cite{fagerberg2012innovation, zhao2015ontology}. For instance, in a single-agent environment, such as an autonomous vehicle navigating a road with obstacles, the first step is to build a knowledge base. This knowledge base includes information on traffic regulations, the environmental map, operation limits of the vehicle, and any restrictions, all organized using appropriate knowledge representation techniques. In addition to the knowledge base, the autonomous agent must adhere to its mission-specific objective, defined as a reach-avoid specification that directs the actions of the agents. The integration of this mission-specific objective, the knowledge base, and sensory data creates a \emph{composite specification} that directs the behavior of the agent in real time. While the reach-avoid specification and knowledge base remain fixed throughout a particular mission, this composite specification can be updated as new sensory data is received, allowing the agent to respond effectively to its environment by updating its control policy.

Knowledge representation techniques have been extensively studied in the existing literature, with methods like predicate logic, frames, semantic networks, and production rules commonly employed~\cite{li2018survey, vickery1986knowledge}. Foundational work has also explored logical, philosophical, and computational aspects~\cite{sowa1999knowledge} of knowledge representation. More recently, research has focused on description logics (DLs) as a framework for structured knowledge representation and reasoning, particularly useful in fields involving complex information and decision-making~\cite{baral2015knowledge}. Within DLs, \emph{attributed concept language} ($\mathcal{ALC}$) plays a central role as a fundamental subset and was initially introduced in~\cite{schmidt1991attributive}. $\mathcal{ALC}$ provides a balance between expressive power and computational manageability~\cite{lawan2019semantic}. However, $\mathcal{ALC}$ cannot represent temporal properties because it lacks inherent support for temporal operators. This limitation makes $\mathcal{ALC}$ unsuitable for applications that involve time-varying features~\cite{baader2012ltl,baader2014runtime}. To capture time-dependent behaviors, a combination of $\mathcal{ALC}$ with linear temporal logic ($\mathcal{ALC}$-LTL) is used to create a knowledge framework for autonomous agents, capitalizing on its strengths in efficiently representing structured knowledge~\cite{baader2012ltl}.

\begin{figure}[t]
    \centering
    \includegraphics[width=1\linewidth]{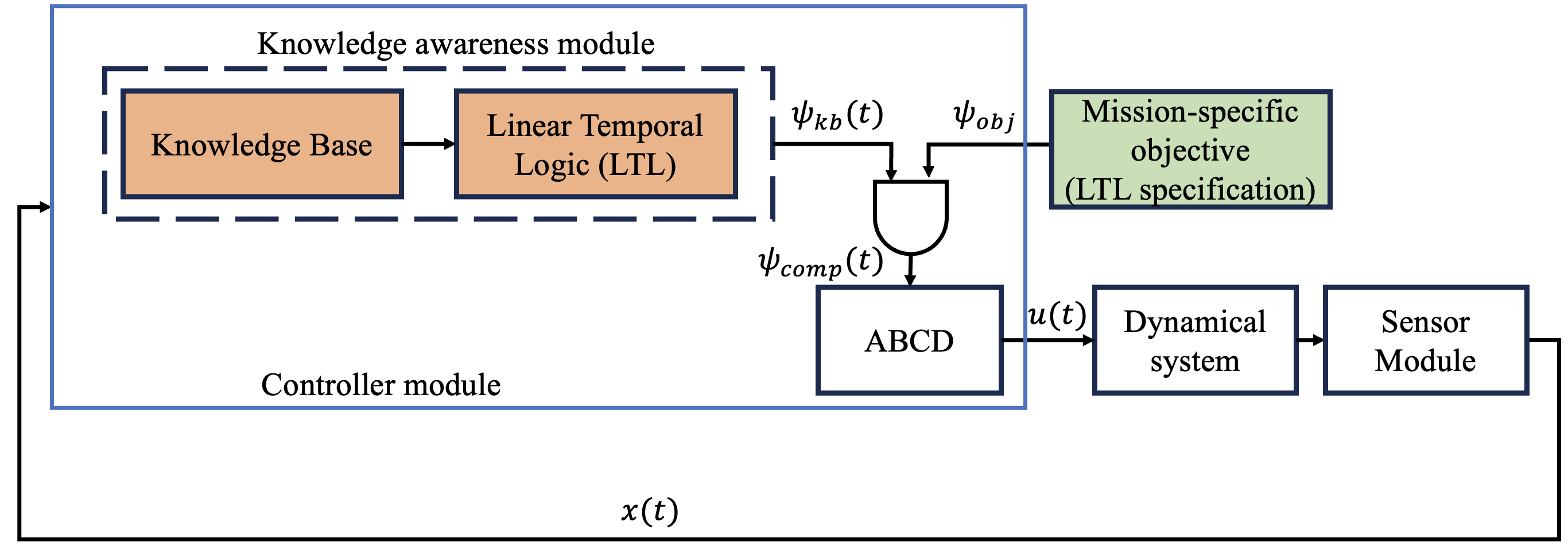}
    \caption{The overall scheme of the proposed approach. The decision-making process relies on three distinct inputs: the LTL formula derived from the knowledge base, the mission-specific objective of the autonomous agent, and the system dynamics.}
    \label{fig:KA_1_agent}
    \vspace{-15pt}
\end{figure}

In this work, we create a knowledge base as a organized repository using $\mathcal{ALC}$-LTL. The pictorial representation of our proposed scheme is shown in Fig.~\ref{fig:KA_1_agent}. At each time point 
$t\in \reals$, during the operation of the autonomous agent, \emph{real-time} sensory data are utilized to extract the relevant subset of information from the knowledge base, forming a temporal specification $\psi_{kb}(t)$. This specification serves as a foundation for introducing knowledge awareness. In addition, the agent has a mission-specific objective $\psi_{obj}$, also represented in LTL. 
These two specifications are then combined to form a composite time-varying specification $\psi_{comp}(t)$ that must be followed by the autonomous agent. Once this composite specification is established, we focus on the second challenge of synthesize a control input that ensures formal guarantees regarding its satisfaction. To achieve this, we consider agents whose dynamics are described using nonlinear set of differential equations with bounded disturbances. Formal controller design for continuous dynamical systems is challenging due to their continuous state space. One promising approach to tackle the formal controller synthesis problem for nonlinear dynamical systems is abstraction-based controller design (ABCD) \cite{Tabuada2009book,belta2017formal,reissig2016feedback,rupak2020ouputfeedback, Majumdar2021, Milad:2022, Majumdar2023}. ABCD schemes construct a finite abstraction of a dynamical system that has continuous state and input spaces, and solve a two-player graph game on the abstraction. Standard ABCD is used in an \emph{offline} manner, that is the controller is synthesized with respect to a given \emph{static specification}. However, this paper studies the use of ABCD in a setting wherein the composite specification $\psi_{comp}(t)$ is subject to change. Therefore, the controller needs to be updated whenever there is a change in the composite specification.

\begin{figure}[t]
    \centering
    \includegraphics[width=0.8\linewidth]{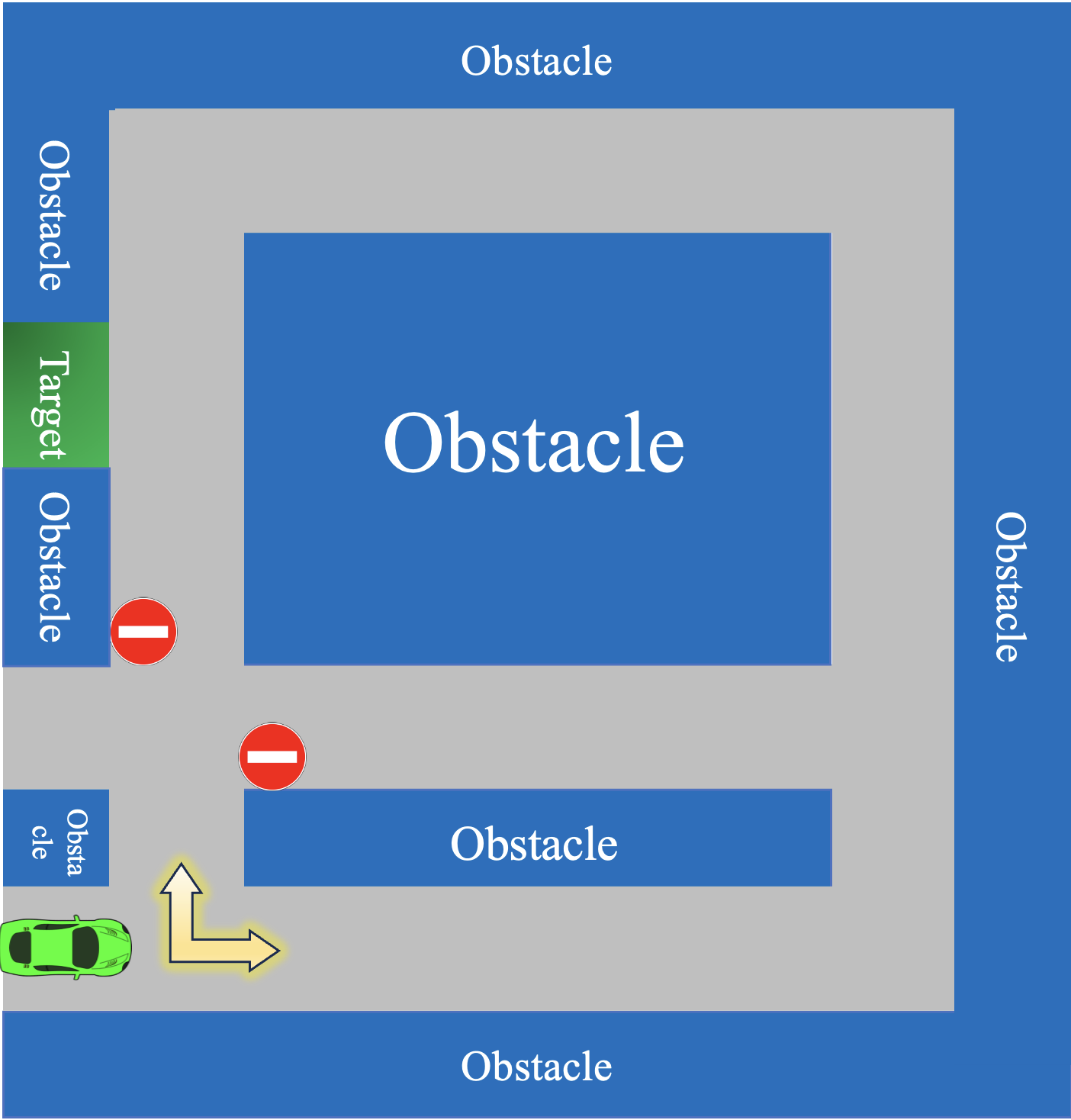}
    \caption{A motivating example of the workspace consisting of an autonomous car, traffic signals, a target destination, and obstacles.}
    \label{fig:motivating_exmp_v2}
    \vspace{-15pt}
\end{figure}

\noindent \textbf{Motivating example.} To illustrate how knowledge awareness can improve decision-making for autonomous agents, we consider a motivating example throughout this paper. This example as shown in Fig.~\ref{fig:motivating_exmp_v2}, serves both as an illustration and eventually as an implemented case study. To this end, we consider an \emph{autonomous vehicle} as an agent that aims to navigate from the \emph{start} (initial) location to the \emph{target} location, while avoiding the \emph{obstacles} and respecting the \emph{traffic rules} set by the traffic signs existing on the map. It is noted that the shortest path passes through the \emph{no-entry} street signs, and hence must be prohibited. The autonomous vehicle begins without the knowledge of the locations of the \emph{no-entry} sign, so it must navigate the map to determine a path from the start to the target, while avoiding obstacles and no-entry areas. 

\noindent \textbf{Contributions.} In this paper, we aim to design a controller integrated with a knowledge base for a nonlinear dynamical system, that provides formal correctness guarantees, and is guided by a mission-specific objective. We translate the knowledge base represented by description logic (DL) statements~\cite{baader2003description} into a linear temporal logic (LTL)~\cite{baier2008principles} formula to serve as a specification. The combination of the mission specific objective and the specification from the knowledge base forms a composite specification, which is time-varying. This composite specification is then integrated with the controller synthesis process, which results in an adaptive control policy. For controller synthesis, we employ an abstraction-based controller design (ABCD) method, which can algorithmically construct a controller with formal correctness guarantees for systems with nonlinear dynamics, bounded disturbances, and general temporal specifications. As a case study, we consider a nonlinear 3-dimensional dynamical car model for which we use ABCD to synthesize a controller that fulfils the composite specification. Thus, our decision-making process relies on three distinct inputs: the LTL formula derived from the knowledge base, the mission-specific objective of the autonomous agent, and the dynamics of the autonomous agent. The overall scheme of the proposed method is shown in Fig.~\ref{fig:KA_1_agent}.

\noindent \textbf{Outline.} The remainder of the paper is structured as follows, section \ref{sec:overview} gives the description of the proposed scheme along with the necessary definitions, section \ref{sec:knowledge_awareness} discusses the knowledge awareness formulation for autonomous driving case study, section \ref{sec:evaluation} presents the considered example, specification, and the formal synthesis method employed, along with the results. Concluding remarks with future directions are provided in section \ref{sec:conclusion}.
\section{Proposed Scheme}\label{sec:overview}
In this section, we outline the various components of our proposed scheme (see Fig.~\ref{fig:KA_1_agent}). We begin by discussing \emph{static} knowledge bases, which are represented using $\mathcal{ALC}$, a fundamental subset of DL used for knowledge representation. Next, we explain how to enhance these knowledge bases with awareness to capture temporal specifications. We then describe the role of the sensor module and the objective within our framework. Finally, we introduce the dynamical systems and the abstraction-based controller design approach.
\subsection{Knowledge base}
As shown in Fig.~\ref{fig:KA_1_agent}, we address the controller synthesis problem where the available information and constraints are stored in a knowledge base. This information can include detailed environmental data, such as traffic rules, obstacle features, road conditions, and priority rules that support decision-making in dynamic settings. As previously mentioned, $\mathcal{ALC}$--a core subset of Description Logic (DL)--is widely used for representing knowledge bases across various fields, in this subsection, we introduce the fundamentals of $\mathcal{ALC}$, including the formal definitions of the symbols it employs.
\begin{definition}[$\mathcal{ALC}$ \emph{symbols}~\cite{kg-book}]
    We define symbols of $\mathcal{ALC}$, $S_{\mathcal{ALC}}\coloneqq B \bigcup N_A \bigcup N_R \bigcup N_I \bigcup Q$, where $B$ represents the \emph{structural symbols} $\{\mlq(\mrq , \mlq)\mrq, \mlq.\mrq\}$, $N_A$, $N_R$, and $N_I$ represent  countably-infinite sets of \emph{atomic concepts},  \emph{atomic roles}, and \emph{instances} respectively, $Q$ represents \emph{logical symbols} $\top, \bot, \neg, \sqcap, \sqcup, \forall, \exists, \sqsubseteq, \equiv$. 
\end{definition}
Intuitively, \emph{instances} refer to elements of the world, where in the context of a control system, the world corresponds to the state space. Further, an \emph{atomic concept} defines a unary relation over the set of instances, while an \emph{atomic role} defines a binary relation over pairs of instances. The logical symbols along with atomic concepts and roles are utilized to create \emph{complex concepts} or simply $\mathcal{ALC}$-\emph{concepts}. In this paper, \emph{atomic concepts} ($A$) are written in uppercase letters, \emph{roles} ($r$) and \emph{instances} ($a$ or $b$) are written in lowercase letters. Furthermore, \emph{complex concepts} are described using $C$ and $D$.  
\begin{definition}[$\mathcal{ALC}$ \emph{concepts}~\cite{baader2008description}]
The set of $\mathcal{ALC}$-concept is the smallest set satisfying the following conditions:
\begin{enumerate}
    \item $\top$, $\perp$, and every atomic concept name $A\in N_A$ are considered $\mathcal{ALC}$-concept.
    \item If $C$ and $D$ are $\mathcal{ALC}$-concept and $r\in N_R$, then $C\sqcap D$, $C \sqcup D$, $\neg C$, $\forall r.C$, and $\exists r.C$ are also $\mathcal{ALC}$-concept.   
\end{enumerate}
\end{definition}
We can describe the different kinds of $\mathcal{ALC}$-\emph{concept} using the following grammar definition~\cite{glimm2019classical}:
\begin{equation}
C, D \coloneqq  A \ | \top | \perp | \neg C | C\sqcap D | C\sqcup D |\exists r.C|\forall r.C.\label{ref:eqn:concept_def}
\end{equation}
Equation~(\ref{ref:eqn:concept_def}) implies that the set of $\mathcal{ALC}$ concepts $C$ and $D$ are recursively constructed starting from an \emph{atomic concept} $A$, \emph{top concept} $\top$, \emph{bottom concept} $\perp$, by applying \emph{negation} $\neg C$, \emph{conjunction} $C\sqcap D$, disjunction $C\sqcup D$, \emph{existential restriction} $\exists r.C$, and \emph{universal restriction} $\forall r.C$. Here, $\top$ represents the set of all objects, $\perp$ is the empty set of objects, $\neg C$ represent all objects that are not in $C$, $C\sqcap D$ represent the set of common objects in $C$ and $D$, $C\sqcup D$ represents all objects in $C$ and $D$, $\exists r.C$ represents all object that are related by $r$ to some object in $C$, $\forall r.C$ represents all objects that are related by $r$ to only objects in $C$. 

Once $\mathcal{ALC}$-concept are formed, they can be utilized to describe various characteristics by composing $\mathcal{ALC}$-\emph{axioms} (formulas $\mathcal{F}_\mathcal{ALC}$)~\cite{kg-book,glimm2019classical}. There are four different types of $\mathcal{ALC}$-\emph{axioms}: 1. \emph{Concept inclusion} ($C\sqsubseteq  D$); 
2. \emph{Concept equivalence} ($C \equiv D$); 
3. \emph{Concept assertion} ($a:~C$);  
4. \emph{Role assertion} ($(a, b):~r$). Axioms are typically assembled into collections to create \emph{knowledge base} (or \emph{ontologies}). A DL $\mathcal{ALC}$ \emph{knowledge base (KB)} comprises two components: a \emph{terminological} part, known as the \emph{TBox}, and an \emph{assertional} part, known as the \emph{ABox}. Each part contains a collection of axioms. Now, we are able to formally define knowledge bases represented by $\mathcal{ALC}$.  
\begin{definition}[\emph{Knowledge base (KB)}~\cite{baader2003description,baader2008description}]
A knowledge base (KB) is represented as a pair $\mathcal{K}=(\mathcal{T} , \mathcal{A})$, where $\mathcal{T}$ denotes a TBox and $\mathcal{A}$ represents an ABox. A finite set of \emph{concept inclusion} ($C\sqsubseteq  D$) and \emph{concept equivalence} ($C\equiv  D$) is referred to as a TBox and denoted as $\mathcal{T}$. A finite collection of concept assertion ($a:~C$) and role assertion ($(a, b):~r$) is termed an ABox and denoted as $\mathcal{A}$. 
\end{definition}

The following definition provides the formal characterization of the $\mathcal{ALC}$ interpretation.
\begin{definition}[$\mathcal{ALC}$ \emph{interpretation}~\cite{baader2003description,baader2008description}]
An interpretation $\mathcal{I}_{\mathcal{ALC}}=(\Delta^\mathcal{I},f^\mathcal{I})$ comprises a nonempty set $\Delta^\mathcal{I}$ referred to as the domain of $\mathcal{I}_{\mathcal{ALC}}$, and a function $f^\mathcal{I}$ that assigns to each atomic concept $A\in N_A$ a set $A^\mathcal{I}\subseteq \Delta^\mathcal{I}$, to each atomic role $r\in N_R$ a binary relation $\Delta^\mathcal{I} \times \Delta^\mathcal{I}$ and to each instance $a\in N_I$ an element $a^\mathcal{I}\in \Delta^\mathcal{I}$. Considering $C$ and $D$ as $\mathcal{ALC}$ concepts, the semantics are defined as

\begin{align*}
    \top^\mathcal{I}=&\Delta{\mathcal{I}};~\perp^\mathcal{I}=\emptyset;~(C \sqcap D)^\mathcal{I}=C^\mathcal{I}\cap D^\mathcal{I};\\
    ~(C \sqcup D)^\mathcal{I}=&C^\mathcal{I}\cup D^\mathcal{I};~\neg C=\Delta^\mathcal{I}\setminus C^\mathcal{I}; \\
    (\exists r.C)=&\{x \in \Delta^\mathcal{I}~\mid~\text{There is some}~y \in \Delta^\mathcal{I}~\\ 
    &\text{with}~ \langle x,y \rangle\in r^\mathcal{I}~\text{and}~ y \in C^\mathcal{I}\};\\
    (\forall r.C)=&\{x \in \Delta^\mathcal{I}~\mid~\text{For all}~y \in \Delta^\mathcal{I}~ \text{if}~ \langle x,y \rangle\in r^\mathcal{I}\\
    &~\text{then}~ y \in C^\mathcal{I}\} .
\end{align*}
\end{definition}
We define $C^\mathcal{I}(r^\mathcal{I})$ as the extension of the concept $C$ ( role name $r$) within the interpretation $\mathcal{I}$. If $a\in C^\mathcal{I}$, then $a$ is an instance of $C$ in $\mathcal{I}$. Finally, given a knowledge base $\mathcal{K}$ represented through $\mathcal{ALC}$ axioms, the following definition provides a formal characterization of the satisfaction relation.
\begin{definition}[$\mathcal{ALC}$ \emph{Satisfaction relation}~\cite{baader2003description,baader2008description}]
Given a KB $\mathcal{K}=(\mathcal{T}, \mathcal{A})$, $\mathcal{K}$ is called consistent if it has a model. The concept $C$ is called specifiable w.r.t $\mathcal{K}$ if there exists a model of $\mathcal{I}$ of $\mathcal{K}$ with $C^\mathcal{I}\neq \emptyset$. Such interpretation is called a model of $C$ w.r.t. $\mathcal{K}$. Concept $D$ subsumes concept $C$ w.r.t. $\mathcal{K}$ ($\mathcal{K}\models C\sqsubseteq D$) if $C^\mathcal{I}\subseteq D^\mathcal{I}$ for all models $\mathcal{I}$ of $\mathcal{K}$. Concepts $C$ and $D$ are considered equivalent w.r.t. $\mathcal{K}$ ($\mathcal{K}\models C\equiv D$) if they subsume each other w.r.t. $\mathcal{K}$. $a$ is an instance of a concept $C$ w.r.t. $\mathcal{K}$ ($\mathcal{K}\models a: C$) if $a^\mathcal{I} \in C^\mathcal{I}$ holds for all models $\mathcal{I}$ of $\mathcal{K}$. A pair of instances of a concept $(a,b)$ represents an instance of a role name $r$ w.r.t. $\mathcal{K}$ ($\mathcal{K}\models (a,b):r $) if $\langle a^\mathcal{I},b^\mathcal{I}\rangle \in r^\mathcal{I}$ holds for all models $\mathcal{I}$ of $\mathcal{K}$. 
\end{definition}
\begin{definition}[$\mathcal{ALC}$~\cite{kg-book}]
    We can characterize $\mathcal{ALC}$ as a tuple $(\mathcal{S}_{\mathcal{ALC}},\mathcal{F}_\mathcal{ALC},\mathcal{I}_\mathcal{ALC},\models)$, where $\mathcal{S}_{\mathcal{ALC}}$ denotes a set containing the symbols of the logic $\mathcal{ALC}$, $\mathcal{F}_\mathcal{ALC}$ represents a set comprising finite sequences of these symbols, termed as the formulas within the logic $\mathcal{ALC}$, $\mathcal{I}_\mathcal{ALC}$ signifies a set encompassing algebraic structures referred to as interpretations of the logic $\mathcal{ALC}$, and $\models \subseteq \mathcal{I}_{\mathcal{ALC}}\times \mathcal{F}_\mathcal{ALC}$ constitutes a binary relation between interpretations and formulas, designated as the satisfaction relation. The syntax of $\mathcal{ALC}$ is represented in the pair $(\mathcal{S}_{\mathcal{ALC}}, \mathcal{F}_\mathcal{ALC})$, while the semantics of $\mathcal{ALC}$ is represented by the pair $(\mathcal{I}_\mathcal{ALC},\models)$.
\end{definition}
\subsection{Enhancing knowledge base with awareness}
A knowledge base that is characterized by $\mathcal{ALC}$ can represent structured, static knowledge but lacks the ability to express time-varying information, which is critical for many dynamic applications~\cite{baader2012ltl}. $\mathcal{ALC}$ can effectively capture concepts, roles, and relationships within a domain but cannot model how these relationships change over time, as it has no inherent support for temporal operators. This limitation makes $\mathcal{ALC}$ unsuitable for applications where time-dependent states are present~\cite{baader2012ltl,baader2014runtime}. In this section, we present knowledge bases capable of capturing both static and dynamic information by combining $\mathcal{ALC}$ with LTL. The result of this combination is a \emph{temporal} specification, denoted as $\psi_{kb}$ (see Fig.~\ref{fig:KA_1_agent}). This formulation enables autonomous agents to not only understand their environment but also respond to changes and sequences over time, providing a robust foundation for real-time, temporally-aware decision-making. In the remainder of this section, we first provide a brief introduction to LTL, followed by an explanation of how it can be combined with $\mathcal{ALC}$.

We consider specifications in the form of formulas in \emph{linear temporal logic} (LTL). LTL allows us to express a wide range of temporal specifications that are crucial in many control systems. In contrast to Boolean logic, which focuses on current states, LTL enables reasoning about how system's states evolve over time.  Here, we give a brief introduction to LTL. For detailed syntax and semantics of LTL, we refer to the book by Baier and Katoen \cite{baier2008principles} and references therein.
\begin{definition}[\emph{LTL}]
     An LTL formula $\phi$ can be defined as \begin{equation*}
	\psi :=  \operatorname{true} \,|\, p \,|\, \neg \psi \,|\,\psi_1 \wedge \psi_2 \,|\, \mathord{\bigcirc} \psi \,|\, \psi_1\mathbin{\sf U} \psi_2,
\end{equation*} 
where $p$ is an element of the set of atomic propositions $AP$. Let $\rho=\rho_0,\rho_1,\dots$ be an infinite sequence of elements from $2^{AP}$ and denote $\rho_i=\rho_i,\rho_{i+1},\dots$ for any $i\in \nats$. Then  
the satisfaction relation between $\rho$ and a property $\psi$, expressed in LTL,  is denoted by $\rho\models\psi$. We denote $\rho\models p$ if $\rho_0\in p$.
Furthermore,
$\rho\models \neg \psi$  if $\rho$ does not satisfy $\psi$ and 
$\rho\models \psi_1\wedge\psi_2$ 
if $ \rho\models \psi_1$ and $\rho\models \psi_2$.
For \emph{next} operator, $\rho\models\mathord{\bigcirc}\psi$ holds if $\rho_1\models\psi$.
The temporal until operator $\rho\models \psi_1\mathbin{\sf U}\psi_2$  holds if $ \exists i \in \mathbb{N}:$ $\rho_{i} \models \psi_2, \mbox{and } 
\forall j \in{\mathbb{N}:} 0\leq j<i, \rho_{j}\models \psi_1
$. Disjunction ($\vee$) can be defined by
$ \rho\models \psi_1\vee\psi_2\ \Leftrightarrow  \  \rho\models \neg(\neg\psi_1 \wedge \neg\psi_2)$. The operator $\rho\models\LTLeventually \psi$ is used to denote that the property will eventually happen at some point in the future. The operator $\rho\models\LTLalways \psi$ signifies that $\psi$ must always be true at all times in the future. We also define $\psi_1\rightarrow \psi_2$ with $\neg\psi_1 \vee \psi_2$. 
\end{definition}

Baader et al.~\cite{baader2012ltl} proposed combining (extensions of) $\mathcal{ALC}$ with LTL, allowing temporal operators to be applied to concept descriptions, because the expressiveness of DLs is insufficient to describe temporal patterns. 

\begin{definition}[$\mathcal{ALC}$-LTL]\label{def:alc-ltl}
     $\mathcal{ALC}$-LTL formulas are established through induction as:
    \begin{itemize}
        \item If $\alpha$ represents an $\mathcal{ALC}$-\emph{axiom}, then $\alpha$ is an $\mathcal{ALC}$-LTL formula,
        \item If $\psi_1$ and $\psi_2$ are $\mathcal{ALC}$-LTL formulas, then $\neg \psi$, $\psi_1\wedge\psi_2$, $\psi_1\vee\psi_2$, $\psi_1 \mathsf{U} \psi_2$, and $\bigcirc\psi$ also constitute $\mathcal{ALC}$-LTL formulas.
    \end{itemize}
\end{definition}
We can easily extend the $\mathcal{ALC}$ interpretations to the temporal axioms existing in $\mathcal{ALC}$-LTL using the semantics of LTL. 

\noindent\textbf{Integrating the sensory data.} We note that, in general, knowledge base may include a huge set of information, some of which may only become relevant in particular circumstances. The sensor module in Fig.~\ref{fig:KA_1_agent} acts as both an information-gathering and assessment component, capturing the current state of the system, observing environment, and assessing \emph{concept assertions} and \emph{role assertions}. Sensor measurements facilitate knowledge awareness by selecting the \emph{relevant} subset of information from the knowledge base. For example, at time point $t\in \reals$, if a traffic sign is detected, the sensor module records both the current state of the autonomous agent and the location of the traffic sign. Additionally, by employing a named relation, such as “proximity,” the distance between these recorded locations can be assessed and enforcing respecting the corresponding traffic regulation is enforced by setting $\psi_{kb}(t)$. The specification $\psi_{kb}(t)$ represents the subset of restrictions set by the knowledge base that align with the sensor measurements. 
\subsection{Mission-specific objective}
The information from the knowledge base ($\psi_{kb}(t)$) provides general, context-independent rules and environmental data, like traffic laws and road conditions. Mission-specific objective, denoted as $\psi_{obj}$ in Fig.~\ref{fig:KA_1_agent}, however, is a task-focused objective unique to the current mission, such as reaching a particular location or avoiding specific areas. Thus, $\psi_{kb}(t)$ offers foundational guidance, while $\psi_{obj}$ define the unique goals of each operation. At every time point $t\in \reals$, the composite specification $\psi_{comp}(t)$ is formed through combination of LTL formulas that correspond to $\psi_{kb}(t)$ and $\psi_{obj}$:
\begin{equation}\label{eq:composite_spec}
    \psi_{comp}(t) \coloneqq \psi_{kb}(t)\wedge \psi_{obj}. 
\end{equation}
Composite specification $\psi_{comp}(t)$ is used to compute an adaptive controller throughout the agent's mission.
\subsection{Systems dynamics}
Thus far, we have discussed the formulation of the specification that the controller synthesis must satisfy. Our proposed formal controller synthesis approach requires system dynamics in addition to the target specification. Dynamical system in combination with the synthesized controller has to fulfill the specification (see Fig.~\ref{fig:KA_1_agent}). We consider the class of continuous-state continuous-time dynamical systems characterized by the tuple $\Sigma = (X, U, W, f)$, where
$X\subset \reals^n$ is the compact state space, $U\subset\reals^m$ is the compact input space, and
$W\subset\reals^n$ is the disturbance space being a bounded set of disturbances. 
The vector field $f: X \times U \rightarrow X$ is such that $f(\cdot, u)$ is locally Lipschitz for all $u\in U$.
The evolution of the state of $\Sigma$ is characterized by the differential inclusion
\begin{equation}
	\label{eq:ODE}
	\dot x(t)\in f(x(t),u(t))+W.
\end{equation}
Given a \emph{sampling time} $\tau>0$, an initial state $x_0\in X$, and a constant input $u\in U$, define the \emph{continuous-time trajectory} $\zeta_{x_0, u}$
of the system on the time interval $[0, \tau]$ as an absolutely continuous function $\zeta_{x_0,u}: [0, \tau] \rightarrow X$ such that $\zeta_{x_0,u}(0) = x_0$, and
$\zeta_{x_0,u}$ satisfies the differential inclusion $\dot{\zeta}_{x_0,u}(t) \in f(\zeta_{x_0,u}(t), u) + W$ for almost all $t\in [0,\tau]$.
Given $\tau$, $x_0$, and $u$, we define $\Sol(x_0, u, \tau)$ as the set of all $x\in X$ such that there is a continuous-time trajectory
$\zeta_{x_0,u}$ with $\zeta(\tau) = x$.
A sequence $x_0,x_1,x_2, \ldots$ is a \emph{time-sampled trajectory} for a continuous control system if $x_0\in X$ and for each $i\geq 0$, we have
$x_{i+1} \in \Sol(x_i, u_i, \tau)$ for some $u_i \in U$.
\subsection{Abstraction-based controller design}
Given an LTL composite specification $\psi_{comp}(t)$ at time $t\in\reals$ (Eq.~\eqref{eq:composite_spec}) and system's dynamics (Eq.~\eqref{eq:ODE}), the aim is to calculate a controller which ensures the fulfillment of the specification. In order to satisfy a temporal specification on the trajectories systems with continuous state space, it is generally needed to over-approximate the dynamics of the system with a finite discrete-time model. Let $\Xb\subset X$ and $\Ub\subset U$ be the finite sets of states and inputs, computed by quantizing the compact state and input spaces $X\subset\reals^n$ and $U\subset \reals^m$ using the  rectangular discretization partitions of size $\etab_x\in\reals^n_{>0}$ and $\etab_u\in\reals^m_{>0}$, respectively. 
A \emph{finite abstraction} associated with the dynamics in Eq.~\eqref{eq:ODE} 
is characterized by the tuple $ \bar\Sigma\colon(\Xb, \Ub, T_F)$, where $T_F\subseteq\Xb\times \Ub \times \Xb$ denotes the system's \emph{transition system}. The transition system $T_F$ is defined such that
\begin{align*}
(\xbb,\ubb,\xbb')\in T_F&\Leftrightarrow\exists (\xb,\ub,\xb')\in \ball_{\frac{\etab_x}{2}}(\xbb)\times \ball_{\frac{\etab_u}{2}}(\ubb)\times \ball_{\frac{\etab_x}{2}}(\xbb')\;\\
&\text{s.t.}\;
\xb'\in\Sol(\xb,\ub,\tau),
\end{align*}
where $\ball_{\boldsymbol\varepsilon}(\boldsymbol c)\coloneqq \set{\xb \in \mathbb{R}^n \mid  |\xb-\boldsymbol c|\leq \boldsymbol\varepsilon}$ denotes the ball with center $c$ and radius $\boldsymbol\varepsilon$ in $\mathbb{R}^n$. 
When the dynamics in Eq.~\eqref{eq:ODE} are known and satisfy the required Lipschitz continuity condition, the finite abstraction can be constructed using the method proposed in \cite{reissig2016feedback}. For systems with unknown dynamics, data-driven schemes for learning finite abstractions can be employed \cite{Milad:2022,Arcak:2021, MAKDESI202149}.

For a finite abstraction $\bar\Sigma=(\Xb,\Ub,T_F)$ with the corresponding sampling time $\tau$, a feedback controller at the $i^{th}$ sampling time is denoted by $K_i\subseteq\Xb\times\Ub$. The set of valid control inputs at every state $\xbb\in\Xb$ is defined as $K_i(\xbb)\coloneq \set{\ubb\in\Ub\mid(\xbb,\ubb)\in K_i}$.
We denote the feedback composition of $\bar\Sigma$ with $K_i$ as $K_i\parallel\bar\Sigma$. For an initial state $\xbb^\ast\in\Xb$, the set of trajectories of $(K_i\parallel\bar\Sigma)(\xbb^\ast)$ is the set of sequences $\xbb_0,\xbb_1,\xbb_2,\dots$, s.t. $\xbb_0=\xbb^\ast$, $\xbb_{i+1}\in T_F(\xbb_i,\ubb_i)$ and $\ubb_i\in K_i(\xbb_i)$ for $i\in \nats$. 
At the $i^{th}$ sampling time, sensor module provides the state vector $x(i\tau)$. This measurement influences the specification extracted from the knowledge base and forms $\psi_{kb}(i\tau)$. 
Subsequently, $\psi_{kb}(i\tau)$ is combined with $\psi_{obj}$ to form the composite specification $\psi_{comp}(i\tau)$ (see Fig.~\ref{fig:KA_1_agent}). At every time point $i$, ABCD approach is run to compute an updated controller $K_i$ which is guaranteed to satisfy the specification $\psi_{comp}(i\tau)$.

\section{Knowledge Awareness for Autonomous Driving}\label{sec:knowledge_awareness}
This section presents the knowledge representation technique considering the motivating example of autonomous vehicle driving in an urban road scenario as shown in Fig.~\ref{fig:motivating_exmp_v2}. It is noted that, although we have used an autonomous vehicle as the agent in our case study, the methodologies described in section~\ref{sec:overview} can be extended to other autonomous agents. The scenario in Fig.~\ref{fig:motivating_exmp_v2} is represented in the following way:
\begin{itemize}
    \item Traffic rules and traffic signs are represented as concepts, which can be atomic or complex. For example, the concept $NoEntrySign$ is an atomic concept indicating the presence of a ``No Entry'' sign at some state on the map.
    \item states within the state space are represented as instances of atomic concepts. For instance, if a ``No Entry'' sign exists at a specific state $\xbb$ in a 3-D state space, this can be denoted as  $\xbb: NoEntrySign$.
    \item \emph{roles} represent binary relationships between instances. They connect two entities, allowing the expression of relationships and properties that link instances. For example, if two states $\xbb$ and $\xbb'$ are close to each other, this proximity relationship is represented as $(\xbb,\xbb')\colon Proximity$.
    \item A knowledge base ($\mathcal{K}$) is defined using these concepts, roles, and instances along with temporal operators, which basically contains some traffic rules in our case.
    \item By integrating the sensory information at time $t\in\reals$, the knowledge base is then translated to $\psi_{kb}(t)$, which is then combined with the mission-specific specification $\psi_{obj}$ to form a composite LTL specification $\psi_{comp}(t)$, which is then used in the controller synthesis process. 
\end{itemize}

\subsection{Setting up the knowledge base} 
As discussed, we have used $\mathcal{ALC}$-LTL in this paper. It is noted that by Definition~\ref{def:alc-ltl}, an $\mathcal{ALC}$ \emph{axiom} can also be extended as an $\mathcal{ALC}$-LTL formula. An example of a traffic scenario as $\mathcal{ALC}$-LTL statement is given below, which explains ``there exists a \emph{NoEntrySign} within the \emph{proximity} of the sensor of the vehicle'':
 \begin{equation}
    \exists Proximity.NoEntrySign.\label{eqn:road_example}
\end{equation}
A \emph{named-location} (state) in the state space will meet the knowledge criteria in equation~(\ref{eqn:road_example}) if there is a \emph{named-location} (state) in the state space with the \emph{atomic concept} $NoEntrySign$ linked to the \emph{named-location} (state) by the \emph{atomic role} $Proximity$. This state data can be obtained from the sensor module of the autonomous car.

As discussed, $\mathcal{ALC}$-concepts are used to construct $\mathcal{ALC}$-axioms. A knowledge base, which consists of $\mathcal{ALC}$-axioms, has two main components: the terminological part (TBox) and the assertional part (ABox). In TBox, relevant concepts and roles within the application domain are described, along with their properties and relationships and the ABox, serves to describe specific scenarios by specifying attributes of individual entities. 
For the example in Fig.~\ref{fig:motivating_exmp_v2}, we consider the set of \emph{atomic concepts} $\mathsf{A} = \set{Target,~Obstacle,~ NoEntrySign}$, where $Target$ denotes the target destination for the vehicle, $Obstacle$ denotes the obstacles that must be avoided, and $NoEntrySign$ denotes the state at which a \emph{NoEntry} sign exists. Further, the \emph{atomic role} $Proximity$ is denoted as
\begin{align*}
&(\xbb,\xbb')\colon Proximity \Leftrightarrow\nonumber\\
&\xbb,\xbb'\in \Xb\wedge
\min_{x\in \ball_{\eta_x}(\xbb), x'\in\ball_{\eta_x}(\xbb')}d(x,x')<D\wedge\nonumber\\
&\max_{x\in \ball_{\eta_x}(\xbb), x'\in\ball_{\eta_x}(\xbb')}(x_1'-x_1)\cos(x_3)+(x_2'-x_2)\sin(x_3)>0,
\end{align*}
where $d\colon X\times X\rightarrow \reals_{>0}$ calculates the distance of two states $\left((x,x')\mapsto \sqrt{(x_1-x_1')^2 +(x_2-x_2')^2}\right)$, and $D\in \reals_{>0}$ is the maximum detectable distance, which is set based on the camera range of the vehicle. Intuitively, the relation $Proximity$ connects two abstract states $\xbb,\xbb'$ iff their corresponding locations are close and the vehicle's direction is such that it is approaching $\xbb'$.
We create the set of $\mathcal{ALC}$-concepts to form $\mathcal{ALC}$ axioms. 
\begin{equation*}
    \{NoEntrySignDetected,\allowbreak NoEntrySignRespected\}.
\end{equation*}
 The $\mathcal{ALC}$ axioms are then used in the Tbox ($\mathcal{T}$) statements as follows: 
\begin{align}
&\mathcal{T}=\{\nonumber\\
\shortintertext{The vehicle is approaching the street with a \emph{NoEntry} sign}
& NoEntrySignDetected \equiv \exists Proximity.NoEntrySign  \nonumber\\
\shortintertext{Always avoid entering a street with a \emph{NoEntry} sign}
&NoEntrySignRespected \equiv\nonumber\\& \LTLalways (NoEntrySignDetected\rightarrow
\LTLalways \neg NoEntrySign)\}.\label{eqn:tbox_example}
\end{align}
Note that, because we used temporal operators in defining the TBox, the knowledge base is of the type $\mathcal{ALC}$-LTL.

\noindent The corresponding ABox statements are as follows:
\begin{align}
&\mathcal{A}=\{\nonumber\\
\shortintertext{Location of \emph{NoEntry} sign}
&\set{\xbb\in \Xb\mid \xbb\colon NoEntrySign}\nonumber\\
\shortintertext{Locations where the \emph{NoEntry} sign is detected}
&\{\xbb\in \Xb\mid \xbb\colon NoEntrySignDetected\}\nonumber\\
\shortintertext{Locations from where the streets with \emph{NoEntry} sign can be avoided}
& \set{\xbb\in \Xb\mid \xbb\colon NoEntrySignRespected}\nonumber\\
\shortintertext{Set of nearby locations}
&\set{(\xbb,\xbb')\in \Xb^2\mid (\xbb,\xbb')\colon Proximity}\}\label{eqn:abox_example_2}.
\end{align}
\noindent The translation of the TBox statement from equation~(\ref{eqn:tbox_example}) is given as:
\begin{align}
&\forall \xbb(NoEntrySignDetected(\xbb)\Leftrightarrow\nonumber\\
&\exists \xbb'.Proximity(\xbb,\xbb') \wedge NoEntrySign(\xbb')) \nonumber\\
&\wedge \forall \xbb(NoEntrySignRespected(\xbb)\Leftrightarrow \nonumber\\
&\exists K \in 2^{\Xb\times \Ub}\text{ s.t. }(K\parallel\bar\Sigma)(\xbb)\subseteq \nonumber\\
&\LTLalways (NoEntrySignDetected\implies\LTLalways \neg NoEntrySign).
\label{eqn:tbox_example_translated_1}
\end{align}

We note that only those axioms of $\mathcal{T}$ which contain temporal operators should be taken into account for the sake of controller synthesis. Therefore, for every $\xb\in\ball_{\boldsymbol\eta_x/2}(\xbb)$, we have
\begin{align}
&\psi_{kb}(t) =NoEntrySignRespected(\xbb),
\label{eqn:tbox_example_translated}
\end{align}

Due to space limitations, the translation of ABox statements is not included in this paper. However, the translation from equation~(\ref{eqn:abox_example_2}) can be obtained by replacing the variables $\xbb$ with the exact locations from the map. The sensor module will record $\xbb$ and then utilize this information for the evaluation of ABox assertions and feedback them to the knowledge base.
\section{Evaluation}\label{sec:evaluation}
To evaluate our approach, we considered the motivating example in Fig.~\ref{fig:motivating_exmp_v2} of autonomous vehicle driving in an urban scenario to illustrate the value of knowledge awareness in the process of synthesizing controllers. 
\subsection{Dynamics}
We consider the following dynamical model for the vehicle:
\begin{align}
\dot{x}_1(t) =& \cos(x_3(t))\nonumber\\
\dot{x}_2(t) =& \sin(x_3(t))\label{eq:car}\\
\dot{x}_3(t) =& u(t)\nonumber
\end{align}
where, the state variables $x_1,~x_2,~x_3$ represent the position of the vehicle in the $2$-dimensional space and the orientation of the vehicle, respectively. Input $u$ represents the rotational speed. 
The state and input spaces are $X = [0, 8]\times[0, 11]\times[-\pi, \pi]$ and $U =  [-2\pi, 2\pi]$, respectively. We chose the state and input space discretization parameters $\eta_x=(0.15,0.15,0.26)$ and $\eta_u=0.26$. The resulted abstraction has cardinalities $|\Xb|= 78725$ and $|\Ub| = 49$. Also we set the sampling time $\tau=0.2$.
\subsection{Formal synthesis}
We note that ABCD is an offline controller synthesis method, meaning that it takes the dynamics and specification as inputs and provides the formally guaranteed controller as output. However, in our example, the specification is not fully known upfront: initially, only the atomic concepts $Target$ and $Obstacle$ are fully known and the composite specification is initialized as follows:
$$\psi_{comp}(0)=\psi_{obj} = \neg Obstacle \LTLuntil Target.$$
Therefore, the controller takes $\psi_{comp}(0)$ in and generates a controller $K_0\colon \Xb \rightarrow \Ub$. If the initial abstract state $\xbb_0$ is within the domain of $K_0$, the synthesis at the first time step is considered successful, and a control input $\ubb_0\in K_0(\xbb_0)$ is applied within the time interval $t\in [0,\tau)$. At every time step $i\in \nats$, the specification is updated based on the current state of the vehicle $\xb(i\tau)\in \ball_{\frac{\etab_x}{2}}(\xbb_i)$ as follows:
$$\psi_{comp}(i\tau) = \psi_{kb}(i\tau)\wedge \psi_{obj}.$$

\begin{figure*}[t!]
     \centering
     \begin{subfigure}[t]{0.3\textwidth}
         \centering
         \includegraphics[width=\textwidth]{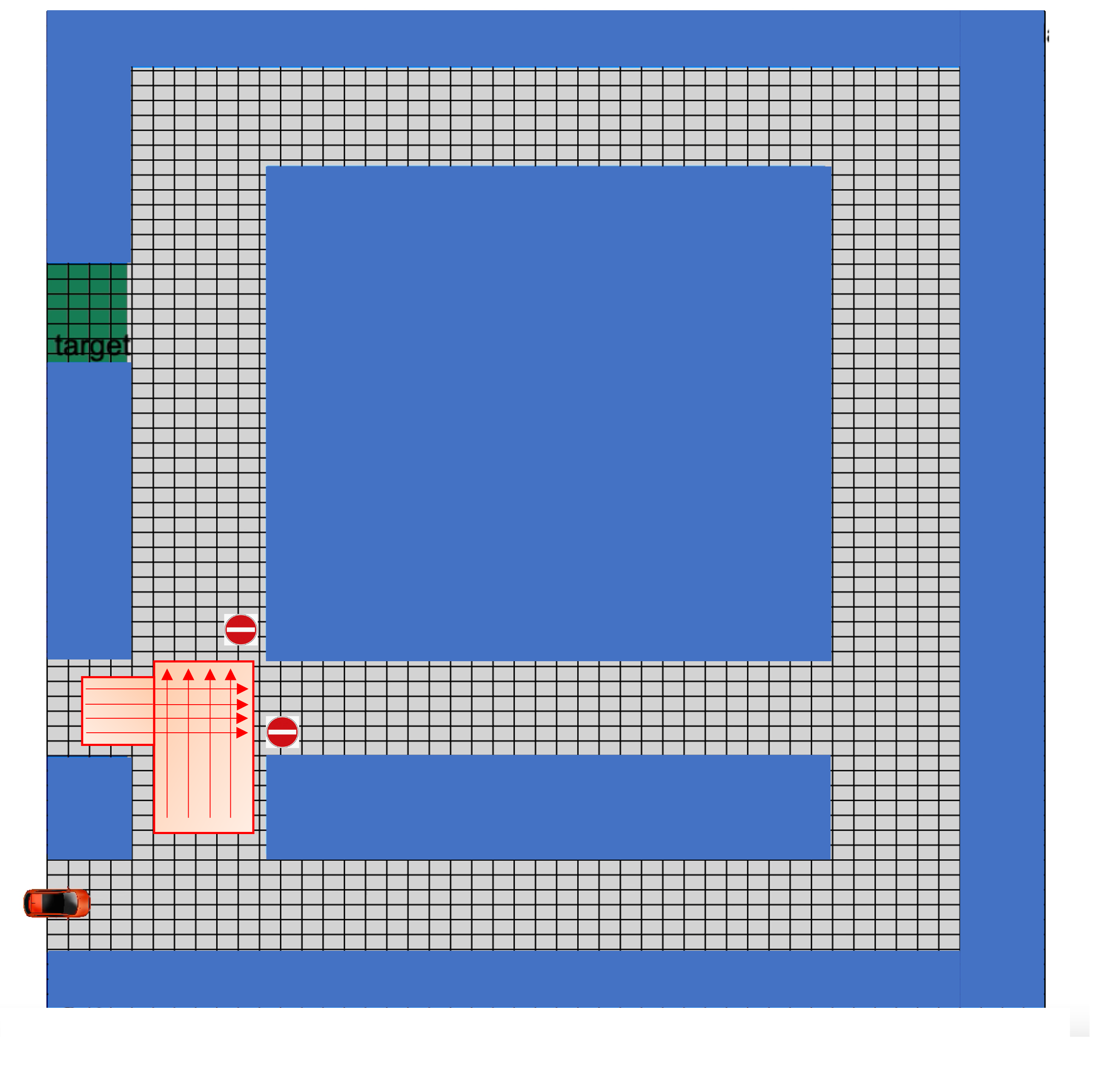}
         \caption{}
         \label{fig:example_ss0}
     \end{subfigure}
     \begin{subfigure}[t]{0.3\textwidth}
         \centering
         \includegraphics[width=\textwidth]{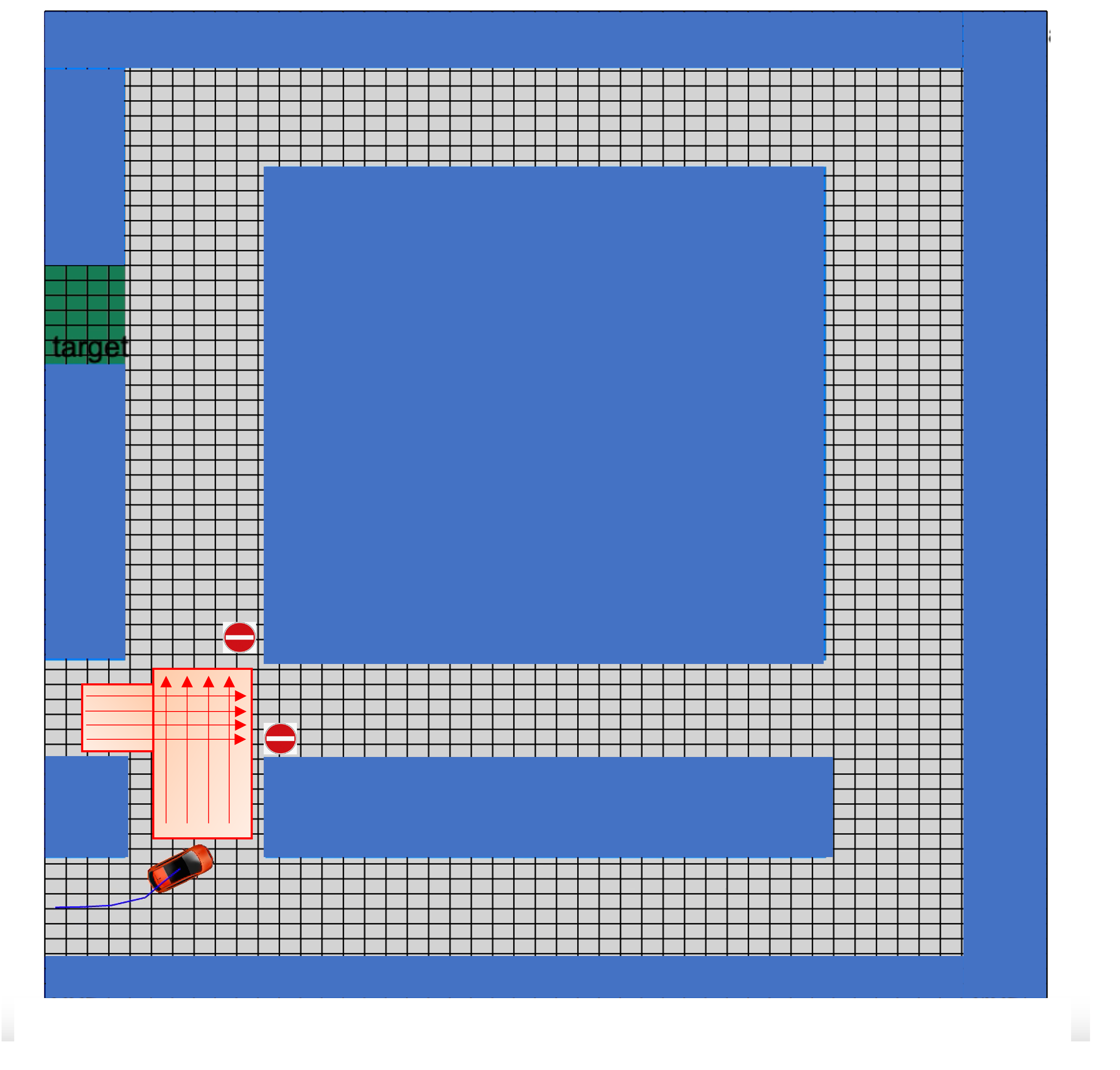}
         \caption{}
         \label{fig:example_ss1}
     \end{subfigure}
     \begin{subfigure}[t]{0.3\textwidth}
         \centering
         \includegraphics[width=\textwidth]{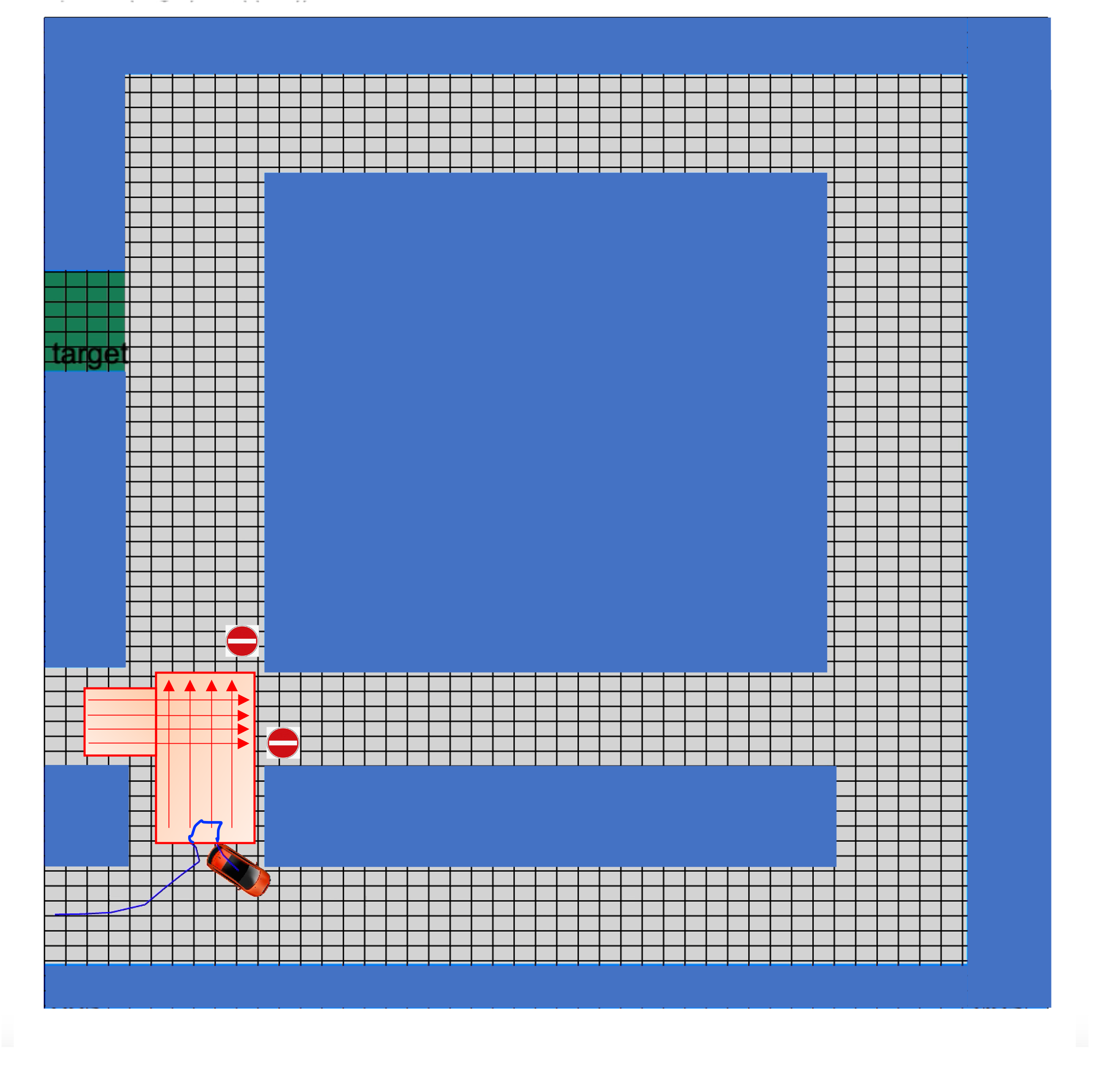}
         \caption{}
         \label{fig:example_ss2}
     \end{subfigure}
     \\
     \begin{subfigure}[t]{0.3\textwidth}
         \centering
         \includegraphics[width=\textwidth]{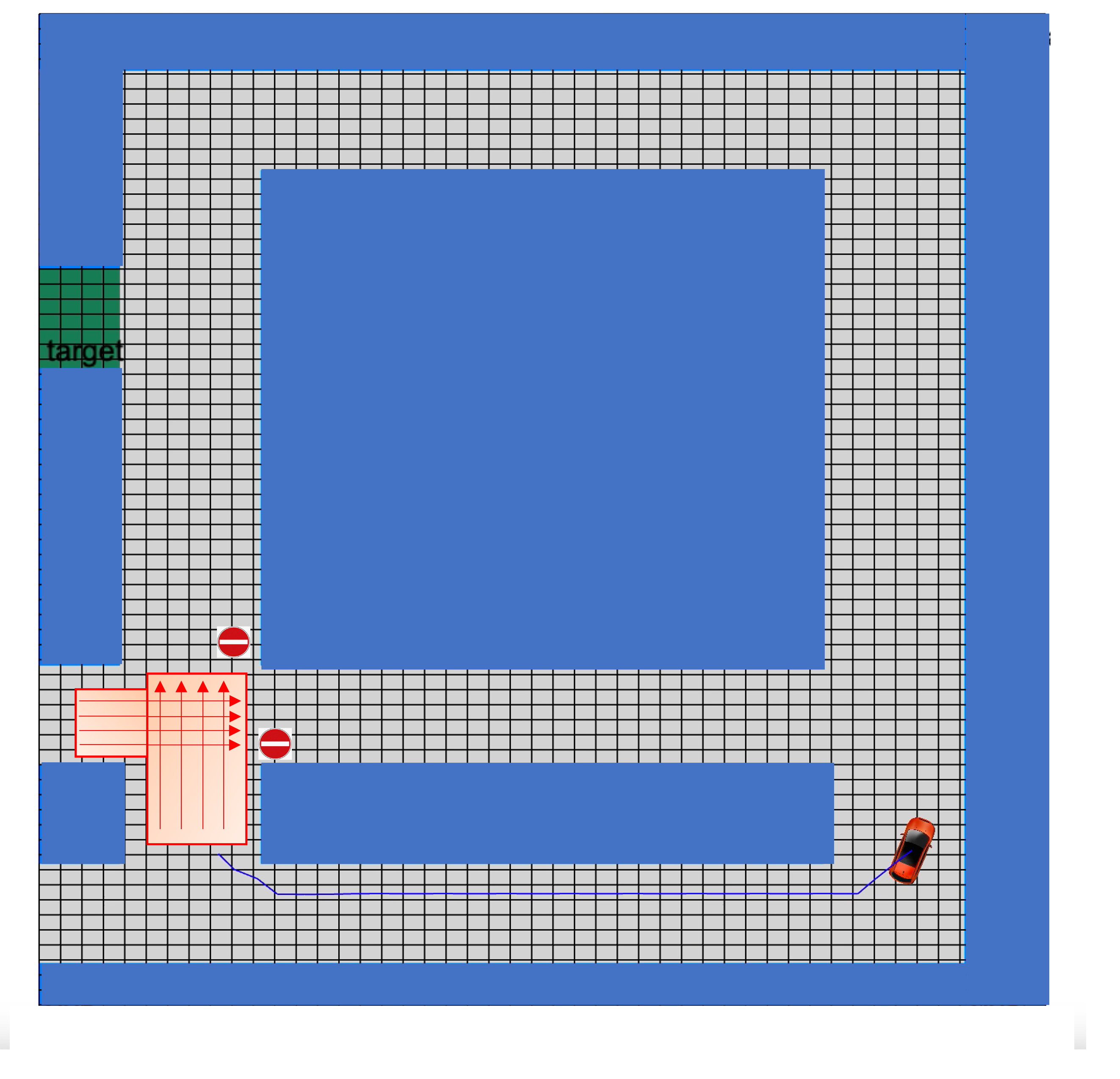}
         \caption{}
         \label{fig:example_ss3}
     \end{subfigure}
     ~
     \begin{subfigure}[t]{0.3\textwidth}
         \centering
         \includegraphics[width=\textwidth]{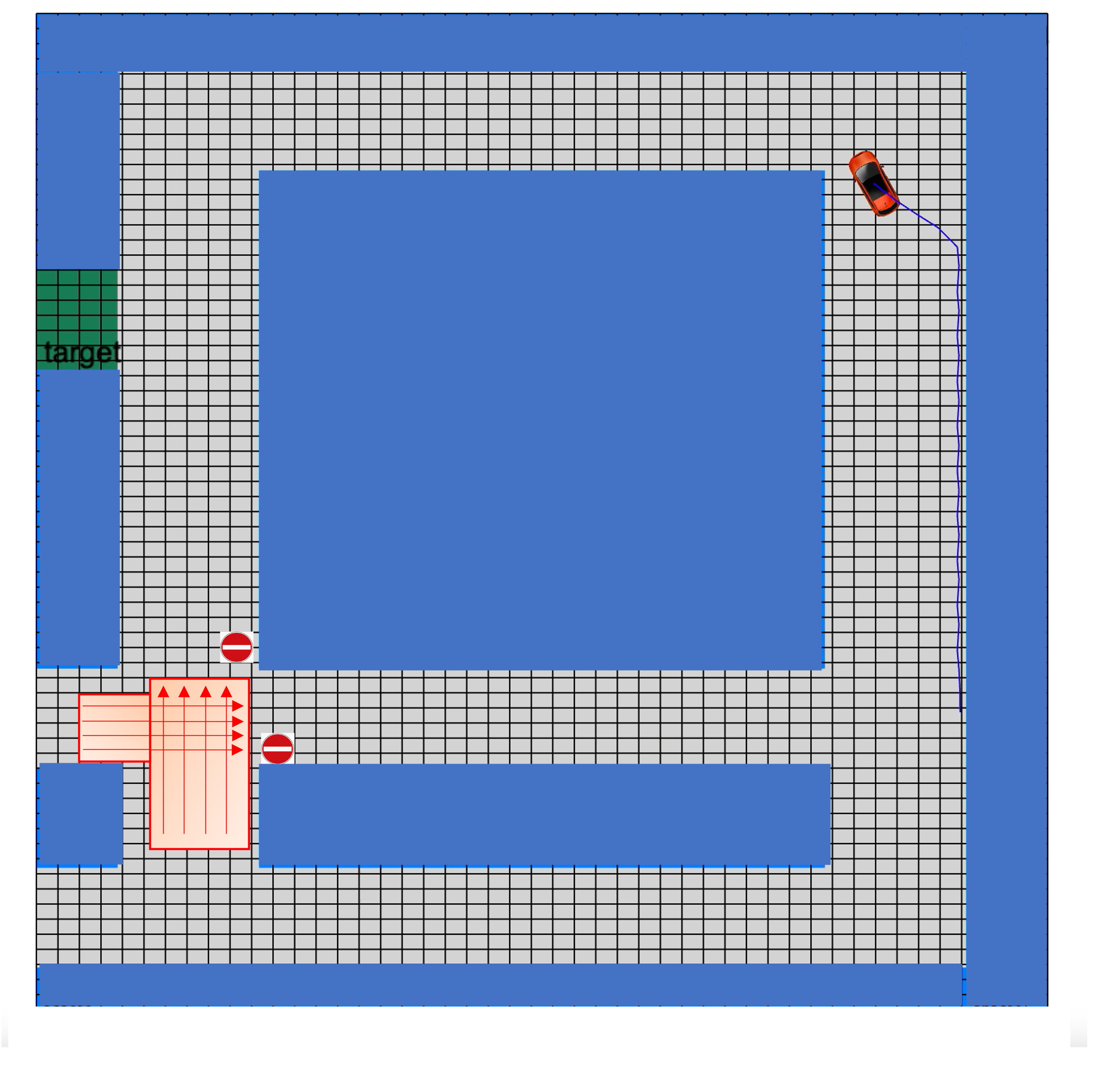}
         \caption{}
         \label{fig:example_ss4}
     \end{subfigure}
    ~
     \begin{subfigure}[t]{0.3\textwidth}
         \centering
         \includegraphics[width=\textwidth]{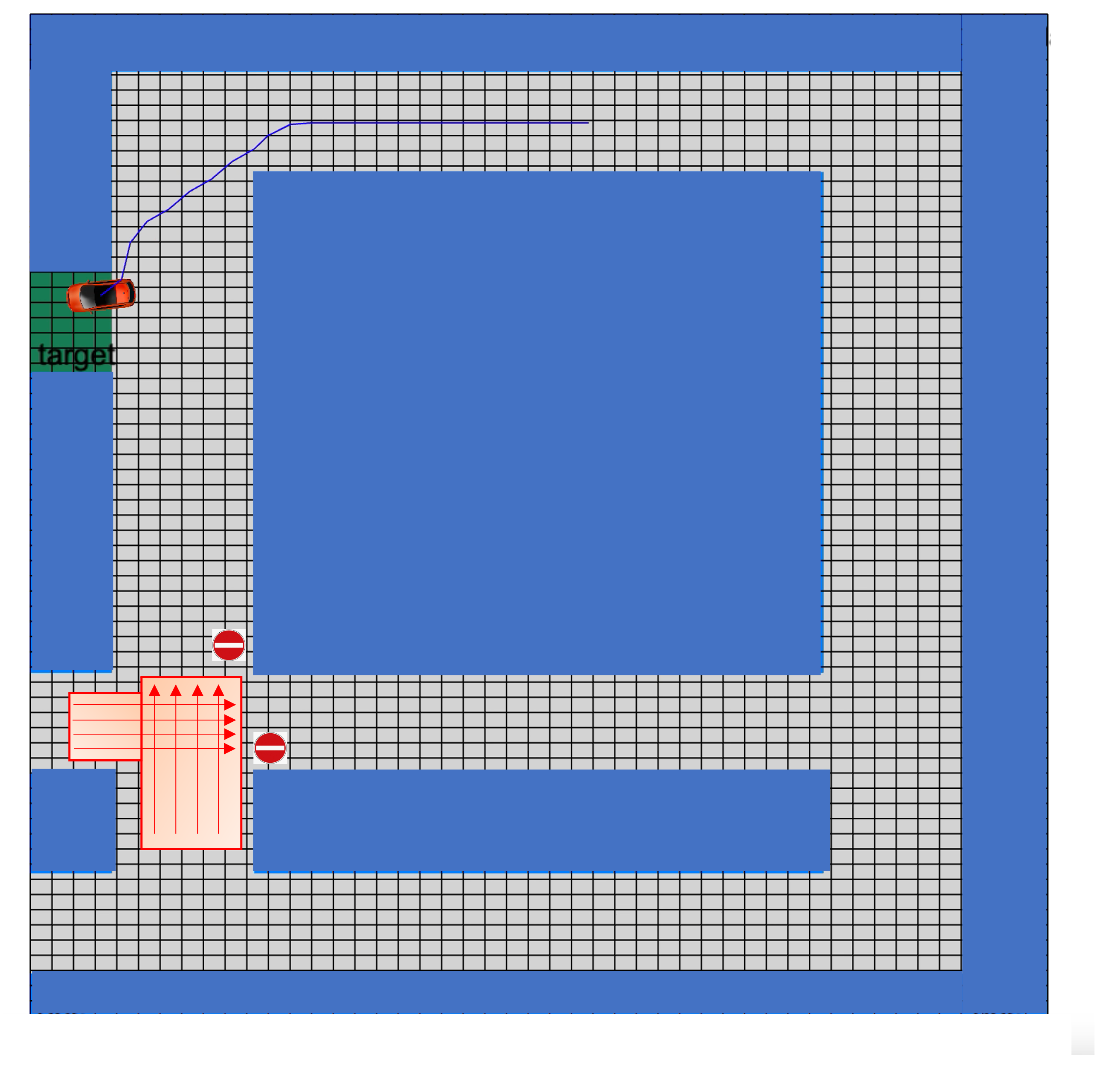}
         \caption{}
         \label{fig:example_ss5}
     \end{subfigure}
        \caption{Depiction of the path followed by the autonomous vehicle: The real-time controller synthesis adheres to both the mission-specific reach-avoid specification and the constraints set by the knowledge base. (a) Initial position of the car; (b) The car makes an immediate left turn towards the target; (c) Due to traffic restrictions, the car turns back to take an alternative route; (d) and (e) The car follows the alternative route towards the target; (f) The car reaches at the target location.}
        \label{fig:toy_examle}
\end{figure*}

Running ABCD, we get a controller $K_i$ that fulfills $\psi_{comp}(i\tau)$ at the $i^{th}$ sampling time. Fig.~\ref{fig:toy_examle} illustrates the path that the vehicle has taken in order to reach its predefined reach-avoid specification, while respecting the restrictions imposed by the knowledge base. The states corresponding to the atomic concepts $Target$, $Obstacle$ and $NoEntrySign$ are specified using colors green and blue, and the traffic sign image for the \emph{NoEntry} sign, respectively. Furthermore, the highlighted area near the \emph{NoEntry} signs denotes the locations and directions from which the \emph{NoEntry} signs are detectable, i.e., the states corresponding to the complex concept $NoEntrySignDetected$. As it can be observed, initially the controller steers the vehicle 
towards the streets which have \emph{NoEntry} sign (Fig.~\ref{fig:example_ss1}). However, once the \emph{NoEntry} sign was detected, the controller steers the vehicle towards an alternative path (Fig.~\ref{fig:example_ss2}-\ref{fig:example_ss4}). Finally, the car reaches its target in Fig~\ref{fig:example_ss5}. The simulation demonstrates the efficiency of the proposed method, where the autonomous vehicle is guided towards the goal with assistance from the knowledge base, while also adhering to traffic regulations.

\section{Conclusions and Future Works}\label{sec:conclusion}
This paper serves as a proof of concept and a foundational step towards designing knowledge-aware controllers for nonlinear dynamical systems. We demonstrated how to translate a structured knowledge base into temporal specifications for controller synthesis while ensuring formal correctness and satisfaction of a mission-specific objective (reach-avoid) for an autonomous agent. Using an abstraction-based design, we developed a controller that meets both knowledge base specifications and mission-specific reach-avoid goals. The proposed controller responds to real-time sensor data, keeping actions aligned with the knowledge base. Validation was conducted using a 3-dimensional nonlinear dynamical car model navigating an urban road scenario with traffic signs and obstacles. Simulation results demonstrate that the controller effectively guides the autonomous agent towards the target, adhering to predefined reach-avoid specifications with the assistance of the knowledge base. Future work will focus on scaling to complex, high-dimensional environments with dynamic obstacles and managing sensor data uncertainty while maintaining formal guarantees. Possible approaches might include reinforcement learning for adaptability and probabilistic methods for uncertainty management.

\bibliographystyle{IEEEtran}
\bibliography{main_ref}
\end{document}